\documentclass[preprint,pra,10pt,twocolumn]{revtex4}%
\usepackage{amsfonts}
\usepackage{amsmath}
\usepackage{amssymb}
\usepackage{graphicx}%
\providecommand{\U}[1]{\protect\rule{.1in}{.1in}}
\providecommand{\U}[1]{\protect\rule{.1in}{.1in}}
\providecommand{\U}[1]{\protect\rule{.1in}{.1in}}
\providecommand{\U}[1]{\protect\rule{.1in}{.1in}}
\providecommand{\U}[1]{\protect\rule{.1in}{.1in}}
\providecommand{\U}[1]{\protect\rule{.1in}{.1in}}
\providecommand{\U}[1]{\protect\rule{.1in}{.1in}}
\providecommand{\U}[1]{\protect\rule{.1in}{.1in}}
\providecommand{\U}[1]{\protect\rule{.1in}{.1in}}
\providecommand{\U}[1]{\protect\rule{.1in}{.1in}}
\providecommand{\U}[1]{\protect\rule{.1in}{.1in}}
\providecommand{\U}[1]{\protect\rule{.1in}{.1in}}
\providecommand{\U}[1]{\protect\rule{.1in}{.1in}}
\providecommand{\U}[1]{\protect\rule{.1in}{.1in}}
\providecommand{\U}[1]{\protect\rule{.1in}{.1in}}
\providecommand{\U}[1]{\protect\rule{.1in}{.1in}}
\providecommand{\U}[1]{\protect\rule{.1in}{.1in}}
\providecommand{\U}[1]{\protect\rule{.1in}{.1in}}
\providecommand{\U}[1]{\protect\rule{.1in}{.1in}}
\providecommand{\U}[1]{\protect\rule{.1in}{.1in}}

\begin{document}
\title{Landau-Ginzburg Perspective of Finite-Temperature Phase Diagrams of a
Two-Component Fermi-Bose Mixture}
\author{Michael Fodor and Hong Y. Ling }
\affiliation{Department of Physics and Astronomy, Rowan University, New Jersey 08028-1700}

\begin{abstract}
We consider a mixture of two-component Fermi and (one-component) bose gases
under the repulsive Bose-Fermi and attractive Fermi-Fermi interaction. We
perform a systematic study of the finite-temperature phase diagrams in the
chemical potential space, identifying, using the Landau-Ginzburg theory, the
features generic to the phase diagrams within the validity of our model. We
apply the theory to explore the physics of correlated BCS pairing among
fermions in a tightly confined trap surrounded by a large BEC gas.


\end{abstract}
\maketitle

\section{Introduction}

Fermions, constrained by the Pauli exclusion principle, behave very
differently from bosons, to which such a principle does not apply. \ In a many
particle system, the former tend to stay away from each other, while, on the
contrary, the latter tend to be gregarious. \ In the most fundamental level,
fermions (leptons and quarks) are the building block for all the matter with
mass, while bosons serve as the mediator for all the fundamental forces in
nature. \ In spite of this vast difference, when mixed together at
temperatures so low that the de Broglie wavelength of the particles becomes
comparable to or even longer than the interparticle spacing, fermions and
bosons can conspire to create fascinating quantum effects at the macroscopic
scale that are of fundamental interest across a broad spectrum of physics,
especially within the disciplines of condensed matter and nuclear\ physics.
\ As most substances in nature solidify before the temperature could reach the
regime where the macroscopic quantum nature of liquid (or gas) can be
manifested, the combination of liquid isotopes between $^{3}$He and $^{4}$He
\cite{bardeen66} has remained the only laboratory accessible system until
recently when the rapid technological advancement in cooling and trapping of
neutral atoms has completely turned the situation around. \ Not only has it
resulted\ in a dramatic proliferation of such systems, including $^{6}$Li -
$^{7}$Li \cite{hulet01,schrek01}, $^{6}$Li - $^{23}$Na \cite{hadzibabic02},
$^{87}$Rb -$^{40}$K\cite{ferrari02,inguscio02,jin04,simoni06}, $^{6}$Li -
$^{87}$Rb \cite{courteille08}, but more significantly, with ultracold atomic
gases, important parameters, including the interaction between particles of
same or different species, can be tuned precisely
\cite{ketterle04,jin04,simoni06,courteille08}, allowing the physics of
Fermi-Bose mixture to be investigated in a well controlled manner, in regimes
possibly well beyond the reach by traditional solid state systems. \ \ 

Recently, by changing the Feshbach detuning across a certain critical point at
which all the minority atoms pair up with the majority ones to form a
molecular Bose condensate, the group at MIT \cite{ketterle08} has successfully
created, from a two-component Fermi mixture with population imbalance, a
quantum gas where Bose molecules are mixed with spin-polarized (unpaired
majority) fermions. \ In addition to confirming an earlier theoretical
prediction of the existence of a transition from full miscibility to phase
separation \cite{viverit00,molmer98}, using the same system, they were able to
determine the effective dimer-fermion scattering length within a reasonable
agreement with the prediction made more than 50 years ago \cite{skorniakov56}
but never verified experimentally, once again demonstrating that the ultracold
atom system provides an excellent experimental platform for testing theories.

Inspired by this work, instead of one Fermi state as in a single-component
Fermi-Bose mixture in Ref. \cite{ketterle08}, we consider a two-component
Fermi-Bose mixture involving a hyperfine state $\left\vert b\right\rangle $ of
a bosonic atom with mass $m_{B}$ and \ two equally populated hyperfine states:
spin up $\left\vert \uparrow\right\rangle $ and spin down $\left\vert
\downarrow\right\rangle $ of a fermionic atom with mass $m_{F}$. \ The latter
Fermi system when equipped with Feshbach resonance has been the main source of
inspiration for much recent excitement in the forefront of ultracold atomic
physics, due chiefly to the vital role it plays in the study of crossover from
Bose-Einstein condensation (BEC) of tightly bound atom pairs to\ nonlocal
Bardeen-Cooper-Schrieffer (BCS) atom pairs. \ Thus, mixtures of bosons with
such Fermi systems shall be widely accessible to experiments.

The low temperature physics of the two-component model under consideration is
dominated by s-wave collisions, which, given that the Pauli exclusion
principle prohibits s-wave scattering between identical fermions, are
parameterized with three scattering lengths:\ $a_{BB}$, $a_{FF}$, and $a_{BF}%
$, describing, respectively, s-wave scattering between two bosons, between two
fermions of opposite spins, and between a boson and a fermion of either spin,
assuming that scattering a boson off a spin-up fermion has the same amplitude
as that off a spin-down fermion. \ In principle, $a_{BB}$\ must be positive
necessitated by the Bose stability in the mixture, while $a_{BF}$ and $a_{FF}$
can take both positive and negative values as they are not subject to similar
constraints. \ Thus, in spite of the restriction on $a_{BB}$, such a model can
still represent drastically different regimes of physics as far as pairings
and instabilities are concerned. \ In the present work, we aim to extend the
physics of phase and phase separation at finite temperature from single- to
two-component systems where fermions have the opportunity to form correlated
BCS pairs. \ As a result, we limit our study to systems with repulsive
($a_{BF}>0$) Bose-Fermi and attractive ($a_{FF}<0$) Fermi-Fermi interaction. \ 

\ In Sec. \ref{sec:thermodynamical potential}, we review, within the framework
of mean-field theory, the path integration formulation of the
finite-temperature thermodynamic potential and derive from it the
Landau-Ginzburg expansion. In contrast to the existing works for the
two-component model \cite{Viverit00,Stoof00,pu10}, which have various
purposes, we focus on producing phase diagrams in the space made up of
chemical potentials, which are intensive statistical variables that must
remain invariant among the separated phases and hence uniquely define a phase
separation \cite{radzihovsky08}. This is to be contrasted to spaces where
coordinates are served, for example, by particle number densities, of which
separated phases in a phase separation have different values
\cite{Viverit00,Salasnich07}. \ In Sec. III, we apply the analytical
intuitions derived from the Landau-Ginzburg theory to identify as well as to
clarify the features that are generic to the finite-temperature phase diagrams
within the validity of our model.\ The utility of such phase diagrams can be
most easily appreciated when one wants to map out the particle number
densities for a trapped model where the local density approximation holds.
\ An example will be shown in Sec. \ref{sec:an application}, which illustrates
how the surrounding large BEC affects the physics of pairing among fermions in
a tightly confined trap. \ Finally, we provide a short conclusion and
discussion in Sec. \ref{sec:conclusion}. \ 

\section{Mean-Field Thermodynamic Potential and Landau-Ginzburg Expansion}

\label{sec:thermodynamical potential}

The thermodynamic properties of our model can be described by the partition
function $Z$, which is a functional integral over both the complex fields for
bosons: $\psi_{\mathbf{k},B}\left(  \tau\right)  $ and $\psi_{\mathbf{k}%
,B}^{\ast}\left(  \tau\right)  $ and the Grassman fields for fermions:
$\psi_{\mathbf{k,}\sigma}\left(  \tau\right)  $ and $\bar{\psi}_{\mathbf{k}%
,\sigma}\left(  \tau\right)  $ in the momentum $\left(  \hbar\mathbf{k}%
\right)  $ and imaginary time $\left(  \tau\right)  $ space \cite{negele98}.
\ In this work, we limit our study to the regime of temperature far below
$T_{B}$ $=2\pi\hbar^{2}\left[  n_{B}/\zeta\left(  3/2\right)  \right]
^{2/3}/(m_{B}\,k_{B})$, where $n_{B}$ is the Bose atom number density, $k_{B}$
the Boltzman constant, and $\zeta\left(  x\right)  $ the Riemann-Zeta
function. \ In this limit, bosons are virtually all condensed to the zero
momentum mode, and the standard symmetry breaking ansatz, $\psi_{\mathbf{k}%
,B}\left(  \tau\right)  =\psi_{\mathbf{0},B}+\phi_{\mathbf{k}\neq0,B}\left(
\tau\right)  $, is therefore applicable, where $\psi_{\mathbf{0},B}$ is a
classical field (not part of the path integration variable) for condensed
particles, and $\phi_{\mathbf{k},B}\left(  \tau\right)  $ is a bosonic field
for condensate excitations or simply phonons for a dilute gas where the usual
Bogoliubov approximation holds \cite{Stoof00}. \ Further, for pedagogical
reason, we opt to first ignore all the phonon degrees of freedom, justifying,
however, in the end of the paper, that the ensuing formalism, when
appropriately modified, can be applied for a large class of Fermi-Bose
mixtures where the effects of phonons are included. \ 

The grand partition function $Z$, at temperature $T\left(  \equiv1/k_{B}%
\beta\right)  $, boson chemical potential $\mu_{B}$, and fermion chemical
potential $\mu_{F}$, then reads $Z=\int D\left[  \bar{\psi},\psi\right]
\exp\left(  -S[n_{B},\bar{\psi},\psi]/\hbar\right)  $, where $n_{B}=\left\vert
\psi_{\mathbf{0},B}\right\vert ^{2}/V$ $\ $is the boson number density, $V$
the total volume, $S$ the action given by%
\begin{align}
S  &  =\hbar\beta V\left(  \frac{g_{BB}}{2}n_{B}^{2}-\mu_{B}n_{B}\right)
+\nonumber\label{S}\\
&  \int_{0}^{\hbar\beta}d\tau\left[  \sum_{\mathbf{k},\sigma}\bar{\psi
}_{\mathbf{k},\sigma}\left(  \hbar\frac{\partial}{\partial\tau}+\xi
_{\mathbf{k}}\right)  \psi_{\mathbf{k},\sigma}\right. \nonumber\\
&  \left.  +\frac{U}{V}\sum_{\mathbf{k},\mathbf{k}^{\prime},\mathbf{q}}%
\bar{\psi}_{\mathbf{k},\uparrow}\bar{\psi}_{\mathbf{q}-\mathbf{k},\downarrow
}\psi_{\mathbf{k}^{\prime},\downarrow}\psi_{\mathbf{q}-\mathbf{k}^{\prime
},\uparrow}\right]
\end{align}
In arriving at Eq. (\ref{S}), we have incorporated the interaction between
fermions and bosons into Eq. (\ref{S}) via $\xi_{\mathbf{k}}=\epsilon
_{\mathbf{k}}-\mu$ in terms of the effective fermion chemical potential
$\mu=\mu_{F}-g_{BF}n_{B}$, where $\epsilon_{\mathbf{k}}=\left(  \hbar
k\right)  ^{2}/2m_{F}$ is the kinetic energy of a fermion with a momentum
$\hbar\mathbf{k}$,\ and $g_{BB}=4\pi\hbar^{2}a_{BB}/m_{B}$ and $g_{BF}%
=4\pi\hbar^{2}a_{BF}/m_{BF}$ measure the strengths of respective s-wave
scatterings with $m_{BF}=2m_{B}m_{F}/(m_{B}+m_{F})$, and we have also modeled
the interaction between fermions of opposite spins with the parameter $U$,
which, in the actual calculation, will be replaced in favor of the physical
parameter $U_{0}\left(  =4\pi\hbar^{2}a_{FF}/m_{F}\right)  $ via the standard
renormalization relation $U^{-1}=U_{0}^{-1}-\sum_{\mathbf{k}}\epsilon
_{\mathbf{k}}^{-1}/(2V)$. \ 

\ In anticipation of the BCS pairing as a result of an attractive
fermion-fermion interaction, we follow the standard procedure in which we
first introduce the auxiliary bosonic fields $\Delta$ and $\Delta^{\ast}$
(assuming they are uniform and static) and then apply the Hubbard-Stratonovic
decomposition to change the partition function into $Z=\int D\left[  \bar
{\psi},\psi,\Delta^{\ast},\Delta\right]  \exp\left(  -S^{\prime}[n_{B}%
,\bar{\psi},\psi,\Delta^{\ast},\Delta]/\hbar\right)  $, where the action
\begin{align}
S^{\prime}  &  =\hbar\beta\left[  V\left(  \frac{g_{BB}}{2}n_{B}^{2}-\mu
_{B}n_{B}\right)  -V\frac{\left\vert \Delta\right\vert ^{2}}{U}+\sum
_{\mathbf{k}}\xi_{\mathbf{k}}\right] \nonumber\\
+  &  \sum_{\mathbf{k},i\omega_{n}}\mathbf{\psi}_{\mathbf{k},i\omega_{n}%
}^{\dag}\left(
\begin{array}
[c]{cc}%
-i\hbar\omega_{n}+\xi_{\mathbf{k}}, & \Delta\\
\Delta^{\ast}, & -i\hbar\omega_{n}-\xi_{\mathbf{k}}%
\end{array}
\right)  \mathbf{\psi}_{\mathbf{k},i\omega_{n}},
\end{align}
is expressed in terms of Nambu spinor $\mathbf{\psi}_{\mathbf{k},i\omega_{n}%
}=(\psi_{\mathbf{k,}i\omega_{n},\uparrow},\bar{\psi}_{-\mathbf{k,-}i\omega
_{n},\downarrow})^{T}$, with $\psi_{\mathbf{k,}i\omega_{n},\uparrow\text{ or
}\downarrow}$ being the field components in the imaginary (Matsubara)
frequency $i\omega_{n}$ space.

Finally, by integrating out fermionic fields and carrying out a summation over
the Matsubara frequency, we arrive, within the saddle point approximation, at
the grand thermodynamic potential (density) [$\Omega=-\ln Z/(\beta V)]$
\begin{align}
\Omega &  =\frac{1}{2}g_{BB}n_{B}^{2}-\mu_{B}n_{B}+\Omega_{F}\left\{
=-\frac{\Delta^{2}}{U}\right.  \nonumber\\
&  \left.  +\frac{1}{V}\sum_{\mathbf{k}}\left[  \xi_{\mathbf{k}}%
-E_{\mathbf{k}}+2\beta^{-1}\ln f\left(  -E_{\mathbf{k}}\right)  \right]
\right\}  \label{Omega}%
\end{align}
for a mixture with Bose particle number density $n_{B}$ and Fermi particle
number density
\begin{equation}
n_{F}=\frac{1}{V}\sum_{\mathbf{k}}\left[  1-\frac{\xi_{\mathbf{k}}%
}{E_{\mathbf{k}}}\tanh\frac{\beta E_{\mathbf{k}}}{2}\right]
,\label{fermi number density}%
\end{equation}
where $E_{\mathbf{k}}=\sqrt{\xi_{\mathbf{k}}^{2}+\Delta^{2}}$ is the
quasiparticle energy, and $f\left(  \varepsilon\right)  =1/(e^{\varepsilon
/k_{B}T}+1)$ is the standard Fermi-Dirac distribution. \ Alternatively, we can
write $\Omega_{F}$ defined in Eq. (\ref{Omega}) as a Landau-Ginzburg expansion
in terms of the order parameter $\Delta$ according to
\begin{equation}
\Omega_{F}=\alpha_{0}+\alpha_{2}\Delta^{2}+\frac{\alpha_{4}}{2}\Delta
^{4}\cdots\label{Omega_F}%
\end{equation}
where
\begin{subequations}
\label{a024}%
\begin{align}
\alpha_{0} &  =\frac{2}{V\beta}\sum_{\mathbf{k}}\ln f\left(  -\xi_{\mathbf{k}%
}\right)  ,\\
\alpha_{2} &  =-\frac{1}{U}-\frac{1}{2V}\sum_{\mathbf{k}}\frac{1-2f\left(
\xi_{\mathbf{k}}\right)  }{\xi_{\mathbf{k}}},\\
\alpha_{4} &  =\frac{1}{V}\sum_{\mathbf{k}}\frac{1}{4\xi_{\mathbf{k}}^{2}%
}\times\nonumber\\
&  \left\{  \frac{1-2f\left(  \xi_{\mathbf{k}}\right)  }{\xi_{\mathbf{k}}%
}-2\beta f\left(  \xi_{\mathbf{k}}\right)  \left[  1-f\left(  \xi_{\mathbf{k}%
}\right)  \right]  \right\}  .
\end{align}
Equations (\ref{Omega}) and (\ref{Omega_F}) serve as the foundations for our
studies below.

\section{Phase Diagrams in Chemical Potential Space}

In this section, we aim to gain a systematic understanding of the phase
diagrams at finite temperature. An immediate obstacle to our goal is the
existence of a large number of free parameters in our model. \ In order to
alleviate this difficulty, we introduce a parameter $A=\hbar^{2}\left(
3\pi^{2}\right)  ^{2/3}/2m_{F}$ which has the physical meaning that
$An_{F}^{2/3}$ equals the Fermi energy $\epsilon_{F}$, and adopt a unit system
generalized from Ref. \cite{viverit00}, in which several key parameters are
scaled according to
\end{subequations}
\begin{align*}
\bar{\Omega} &  =\frac{g_{BF}^{10}\Omega}{g_{BB}^{5}A^{6}},\bar{\mu}_{B}%
=\frac{g_{BF}^{5}\mu_{B}}{g_{BB}^{3}A^{3}},\bar{\mu}_{F}=\frac{g_{BF}^{4}%
\mu_{F}}{g_{BB}^{2}A^{3}},\\
\bar{n}_{B} &  =\frac{g_{BF}^{5}n_{B}}{g_{BB}^{2}A^{3}},\bar{n}_{F}%
=\frac{g_{BF}^{6}n_{F}}{g_{BB}^{3}A^{3}},\bar{k}=\frac{g_{BF}^{2}k}{g_{BB}A}.
\end{align*}
Additionally, $\Delta$, $\mu$, $k_{B}T$, and $\epsilon_{\mathbf{k}}$ are
scaled same as $\mu_{F}$, and the interaction between fermions of opposite
spins [$\bar{U}_{0}=U_{0}/(g_{BF}^{2}/g_{BB})$] is measured relative to
$g_{BF}^{2}/g_{BB}$ - the magnitude of induced fermion-fermion interaction due
to the density fluctuation of Bose condensate \cite{viverit00}. \ As can be
seen, the three original free parameters $\left(  g_{BF},g_{BB},U_{0}\right)
$ are now reduced into a single scaled parameter $\bar{U}_{0}$, making this
unit system particularly suitable for studies aimed at identifying features
generic to all the phase diagrams. \ To gain a qualitative understanding of
this unit system, we note that for a mixture of $^{87}$Rb and $^{84}$Rb with
$a_{BB}=100a_{0}$ and $a_{BF}=550a_{0}$ where $a_{0}$ is the Bohr radius
\cite{bohn99}, the unit is 1.61$\times10^{-49}$ J m$^{3}$ for $U_{0}$,
(2.15$\times10^{-30}$ J, 3.84$\times10^{-31}$ J) for $\left(  \mu_{F},\text{
}\mu_{B}\right)  $, (1.34$\times10^{19}$ m$^{-3}$, 7.49$\times10^{19}$
m$^{-3}$) for $\left(  n_{F}\text{, }n_{B}\right)  $, and 156~$n$K for $T$.
\ (However, for notational simplicity, same symbols will be used to stand for
the scaled variables throughout the rest of the paper.)

\begin{figure}
[ptb]
\begin{center}
\includegraphics[
height=3.7749in,
width=3.4212in
]%
{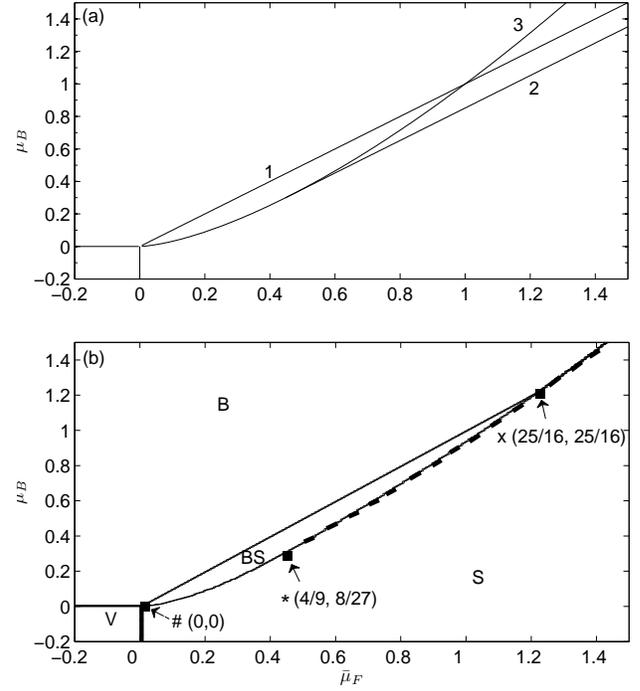}%
\caption{(a) is the illustration designed to aid the understanding of the
zero-temperature phase diagram in (b) where $U_{0}=$-1.126. Units are defined
in the text.}%
\label{Fig:Zero}%
\end{center}
\end{figure}

\begin{figure}
[ptb]
\begin{center}
\includegraphics[
height=3.5544in,
width=3.4757in
]%
{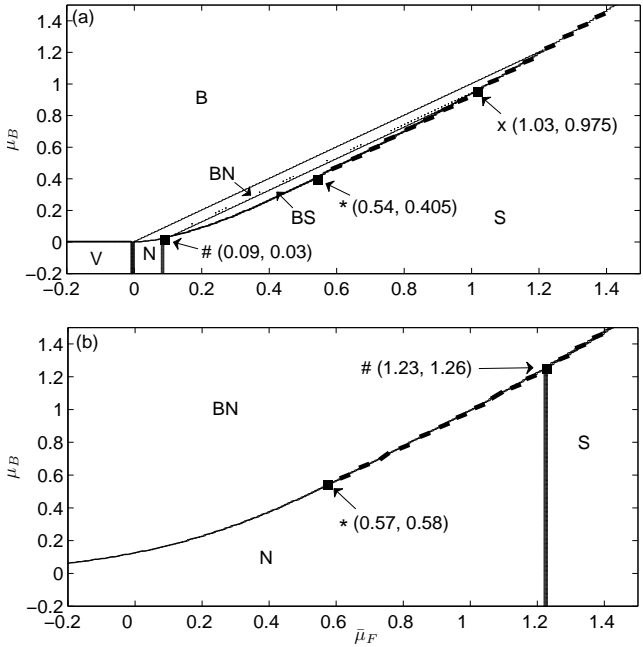}%
\caption{The phase diagram (a) at $T=0.001$ and (b) at $T=0.25$. $U_{0}$ is
same as in Fig. \ref{Fig:Zero}, and units are defined in the text.}%
\label{Fig:Finite}%
\end{center}
\end{figure}

Figures \ref{Fig:Zero}(b) and \ref{Fig:Finite} showcase the features that are
characteristic of both the zero- and the finite-temperature phase diagrams for
a typical system within the validity of our model. They are constructed
numerically by minimizing the potential in Eq. (\ref{Omega}) with respect to
$n_{B}$ and $\Delta$, which amounts to analyzing the saddle point equations
and the relevant Hessian matrix \cite{Stoof00,pu10}, as well as comparing, in
the coexistence regions, the thermodynamic potentials of the possible phases
listed below:

(V) for an empty phase where both $n_{B}=0$ and $n_{F}=0$,

(B) for a pure boson phase where $n_{B}\neq0$ and $n_{F}=0$,

(N) for\ a normal state of Fermi gas where $n_{B}=0$, $\Delta=0$, and
$n_{F}\neq0$,

(S) for a superfluid Fermi gas where $n_{B}=0$, $\Delta\neq0$, and $n_{F}%
\neq0$,

(BN) for a normal Fermi-Bose mixture where $n_{B}\neq0$, $\Delta=0$, and
$n_{F}\neq0$, and finally

(BS) for a superfluid Fermi-Bose mixture where $n_{B}\neq0$, $\Delta\neq0$,
and $n_{F}\neq0$.%
\ \ \ \ 

In order to capture the physics, making the study of phase diagrams more
illuminating, we complement our numerical study with an analysis founded on
the Landau-Ginzburg theory in the low temperature limit $T<<\mu$. \ In this
limit, $\alpha_{4}$ [Eq. (\ref{alpha4}) below] is always positive, and we only
need to consider the Landau-Ginzburg expansion up to the fourth order in
$\Delta$, \
\begin{align}
\Omega &  =\frac{1}{2}n_{B}^{2}-\mu_{B}n_{B}+\alpha_{0}\left(  \mu,T\right)
\nonumber\\
&  +\alpha_{2}\left(  \mu,T\right)  \Delta^{2}+\frac{\alpha_{4}\left(
\mu,T\right)  }{2}\Delta^{4}, \label{Omega_simple}%
\end{align}
where the integrals for the Landau coefficients in Eqs. (\ref{a024}) can be
evaluated explicitly with the results \cite{lifshitz80}
\begin{subequations}
\label{a024-1}%
\begin{align}
\alpha_{0}  &  =-\mu^{5/2}\left[  \frac{2}{5}+\frac{\pi^{2}}{4}\left(
\frac{T}{\mu}\right)  ^{2}\right]  ,\\
\alpha_{2}  &  =-\frac{1}{U_{0}}+\frac{3}{4}\mu^{1/2}\ln\frac{\pi T}{8\mu
e^{\gamma-2}},\\
\alpha_{4}  &  =\frac{21}{32}\mu^{1/2}\frac{\zeta\left(  3\right)  }{T^{2}},
\label{alpha4}%
\end{align}
with $\gamma\approx0.577$ being the Euler's constant. \ Further, in the same
limit, we can ignore the thermal population in comparison with the population
inside the Fermi sphere so that the Fermi density can be approximated as%
\end{subequations}
\begin{equation}
n_{F}=-\frac{\partial\Omega}{\partial\mu_{F}}=\mu^{3/2}\left[  1+\frac{\pi
^{2}}{8}\left(  \frac{T}{\mu}\right)  ^{2}\right]  \simeq\mu^{3/2}, \label{nF}%
\end{equation}
Equations (\ref{a024-1}) and (\ref{nF}) form the backbone of the analytical
discussions that we carry out below. \ 

Consider first the mixed phase where both $n_{B}$ and $n_{F}$ have finite
values. \ We begin with the saddle-point equation for $n_{B}$%
\begin{equation}
n_{B}-\mu_{B}+\mu^{3/2}=0,
\end{equation}
which, when the use of $\mu=\mu_{F}-n_{B}$ is made, leads to a cubic equation
for$\sqrt{\mu}$%
\begin{equation}
\left(  \sqrt{\mu}\right)  ^{3}-\left(  \sqrt{\mu}\right)  ^{2}+\left(
\mu_{F}-\mu_{B}\right)  =0. \label{cubic}%
\end{equation}
The positive root to Eq. (\ref{cubic}) that lies below 2/3 (or$\sqrt{\mu}$
$<2/3$) represents a local stable mixed phase, where the inequality is derived
by subjecting Eq. (\ref{Omega_simple}) to the local stability criteria that
$\partial^{2}\Omega/\partial n_{B}^{2}$ be positive. \ A simple analysis then
shows that Eq. (\ref{cubic}) supports the local stable mixed phase in the
chemical potential space between the line [labeled as 1 in Fig. \ref{Fig:Zero}%
(a)]:
\begin{equation}
\mu_{B}=\mu_{F}, \label{line 1}%
\end{equation}
and the parallel line [labeled as 2 in Fig. \ref{Fig:Zero}(a)] shifted with
respect to Eq. (\ref{line 1}) by $4/27$:
\begin{equation}
\text{ }\mu_{B}=\mu_{F}-4/27. \label{line 2}%
\end{equation}

Next, consider the pure Bose phase described by the bosonic potential
$\Omega_{B}=n_{B}^{2}/2-\mu_{B}n_{B}$ and the pure Fermi phase described by
the fermionic potential $\Omega_{F}=-\frac{2}{5}\mu_{F}^{5/2}$. \ In the
region that supports the pure Fermi (Bose) phase, the effective chemical
potential for fermions $\mu_{F}-n_{B}$ (bosons $\mu_{B}-n_{F}$) must be
negative. \ As a result, the pure Bose phase exists above the line in Eq.
(\ref{line 1}), and the pure Fermi phase exists below the curve [labeled as 3
in Fig. \ref{Fig:Zero}(a)]
\begin{equation}
\mu_{B}=\mu_{F}^{3/2}. \label{line 3}%
\end{equation}

It is then clear that the pure Fermi phase coexists with the mixed phase in
the region below curve 3 but between lines 1 and 2, and it also overlaps with
the pure Bose phase in the upper right region below curve 3 and above line 1.

So far, nothing has been said regarding the nature of the Fermi gas component.
To answer this question, we go to the saddle point equation for $\Delta^{2}$
\begin{equation}
\Delta^{2}=-\frac{\alpha_{2}\left(  \mu,T\right)  }{\alpha_{4}\left(
\mu,T\right)  }.
\end{equation}
Evidently, because $\alpha_{4}\left(  \mu,T\right)  $ is always positive, only
when $\alpha_{2}\left(  \mu,T\right)  <0$ does BCS order or superfluidity
occur. \ Let $\mu_{F}^{\#}$ be the root to the threshold condition%
\begin{equation}
\alpha_{2}\left(  \mu_{F}^{\#},T\right)  =-\frac{1}{U_{0}}+\frac{3}{4}%
\sqrt{\mu_{F}^{\#}}\left(  \ln\frac{\pi T}{8\mu_{F}^{\#}e^{\gamma-2}}\right)
=0. \label{alpha 2}%
\end{equation}
In the limit where $T$ goes to zero, $\mu_{F}^{\#}$ also goes to zero, and
hence as expected under the attractive Fermi-Fermi interaction, fermions at
$T=0$ either exist as a pure S state or mix with\ bosons to form a mixed BS state.

Finally, by comparing the energies of the phases in each coexistent region, we
change Fig. \ref{Fig:Zero}(a) into the phase diagram in Fig. \ref{Fig:Zero}%
(b), which features a tricritical point (labeled as $\ast$) at which the
transition between BS and S changes from the second- to first-order type, and
a critical point (labeled as $\times$) at which the second-order transition
line between B and BS and the two first-order transition lines between S and,
respectively, BS and B meet. \ The explanations are provided as follows. The
first-order transition line (dotted) between B and S is given by
\begin{equation}
\mu_{B}=\sqrt{4/5}\mu_{F}^{5/4}, \label{first-order 1}%
\end{equation}
which is obtained by setting $\Omega_{B}=\Omega_{F}$. \ The interception
between Eq. (\ref{line 1}) and Eq. (\ref{first-order 1}) defines the critical
point ($\mu_{F}^{\times}=\,\mu_{B}^{\times}=25/16$). The comparison between
$\Omega$ and $\Omega_{F}$ determines the first-order transition between BS and
S, which starts from the critical point $\times$ and ends at the tricritical
point $\ast$ with $\left(  \mu_{F}^{\ast}=4/9,\mu_{B}^{\ast}=8/27\right)  $ at
which $n_{B}=\partial\Omega/\partial n_{B}=\partial^{2}\Omega/\partial
n_{B}^{2}=0$. \ 

Note that for a single-component model with the repulsive Bose-Fermi
interaction, the phase diagram at $T=0$ has the same structure as
Fig.\ref{Fig:Zero}(b) except that S and BS are replaced, respectively, with N
and BN, owing to the absence of pairing mechanism for fermions
\cite{ketterle08}. \ As stressed in the introduction,\ an advantage of the
phase diagram in the chemical potential space is that each phase separation
manifests itself as a unique coexistence curve, making its identification
particularly transparent. \ For example, Fig. \ref{Fig:Zero} (b) automatically
rules out phase separations other than those between a pure Fermi and a pure
Bose phase and between a pure Fermi and a mixed phase. \ Thus, there is no
need to hypothesize and analyze the existence of, for example, a phase
separation between a pure bosonic and a mixed phase or that between two
different mixed phases, as did in works where phase analysis was performed in
space made up of particle number densities \cite{Viverit00,Salasnich07}.%
\ 

At (low) finite temperature [Fig. \ref{Fig:Finite}(a)], the pure Fermi phase
is divided into N and S by the vertical (second-order transition) line
$\mu_{F}=\mu_{F}^{\#}$, where $\mu_{F}^{\#}$ increases with temperature
according to Eq. (\ref{alpha 2}), and the mixed phase is split into BN and BS
by the (second-order transition) line, $\mu_{B}=\mu_{F}+n_{F}^{\#}-\mu
_{F}^{\#},$ which begins at point \# and ends at point $\times$, with
$n_{F}^{\#}$ being the Fermi number density [Eq. (\ref{fermi number density})]
at $\Delta=0$ and $\mu=\mu_{F}^{\#}$. \ As the temperature increases, both the
$N$ and $BN$ region expand while $S$ and $BS$ contracts. \ A further increase
in the temperature so that $\mu_{F}^{\#}$ lies beyond $\mu_{F}^{\ast}$ at the
tricritical point results in a complete elimination of the phase space for the
superfluid Fermi-Bose (BS) mixture as shown in Fig. \ref{Fig:Finite}(b), where
the normal state Fermi-Bose mixture remains as the only mixed phase.

At this point, we recall that our analysis rests upon an assumption that the
term $\pi^{2}T^{2}/8\mu^{2}$ in Eq. (\ref{nF}) be much less than 1, which is
therefore applicable only to situations where the Fermi surface is well
defined (relative to the temperature of interest). At zero temperature, the
Fermi surface is fixed by $\mu=4/9$ for line 2 (and the tricritical point),
but is not defined for line 1 where $\mu=0$; line 1 holds only for $T=0$. \ As
a result, the line dividing BN and BS in Fig. \ref{Fig:Finite}(a) (which
corresponds to line 1 at zero temperature in Fig. \ref{Fig:Zero}) is far more
sensitive to the temperature increase than the first-order transition line
(and the tricritical point). \ 

The line between B and BN requires a special explanation. \ At zero
temperature, due to the Fermi-Dirac distribution, the chemical potential for a
normal state of Fermi gas must be positive, and under such a circumstance, the
pure Bose phase (B) is well-defined - the space above line 1, where the
chemical potential is negative $\mu<0$, is defined as the phase B. \ As
temperature increases, the Fermi surface becomes less sharply defined, and so
does the distinction between B and BN. In fact, at finite temperature there is
nothing to prevent the chemical potential for fermions from becoming negative,
just as for distinguishable particles that follow the Maxwell's distribution.
\ In our study here, the line between B and BN is defined in such a way that
the Fermi density along this line is 10$^{-12}$, a small threshold
artificially introduced to indicate that the region in B is virtually free of
fermions. \ But, we stress that in contrast to the line between BN and BS
across which there is a second-order phase transition due to the emergence of
the order parameter $\Delta$ which breaks the symmetry, there is no real phase
transition for fermions across the line between B and BN. \ 

\section{An Application: Tightly trapped Fermions Embedded in a Large BEC}

\label{sec:an application}

In practice, phase diagrams come in different forms, depending on the
application at hands. \ A complete mapping of the phase diagram including all
the critical points in the chemical potential space as we have done in the
previous section is significant, not only because it provides a roadmap for
constructing phase diagrams in other spaces, but also because it can greatly
facilitate both the identification and the interpretation of the important
features emerging from these diagrams. \ As an application, we consider, in
this section, a double species experiment in which spherically symmetric
harmonic trap potentials, $V_{B,F}\left(  r\right)  $, are tuned to differ in
such a manner that fermions are tightly confined inside a much larger BEC, and
use it as a model to explore how the surrounding bosons affect the physics of
Fermi pairing. \ This model is the two-component analog of the
single-component system experimentally realized by the MIT group in 2002
\cite{hadzibabic02}. It also represents an example in which the traps are
oppositely arranged compared to those adopted to implement the idea of using a
small\ trapped BEC to probe the properties of a large Fermi gas\ component
\cite{pu10,bhongale08}.

In our study below, we assume that the Fermi component, although small
compared to the BEC, is still sufficiently large so that the local-density
approximation is directly applicable. Thus, for fermions, we introduce a local
chemical potential $\mu_{F}\left(  r\right)  =\mu_{F}-V_{F}\left(  r\right)
$, in addition to a global chemical potential $\mu_{F}$, which is fixed by
the\ total particle number. \ In contrast, for bosons, owing to its large
size, to a good approximation, we\ can regard its chemical potential, within
the distance\ scale in the order of the size of the Fermi gas, to be a
constant, $\mu_{B}$, independent of the variation of the radial distance $r$. \ 

\begin{figure}
[ptb]
\begin{center}
\includegraphics[
height=3.5016in,
width=3.4238in
]%
{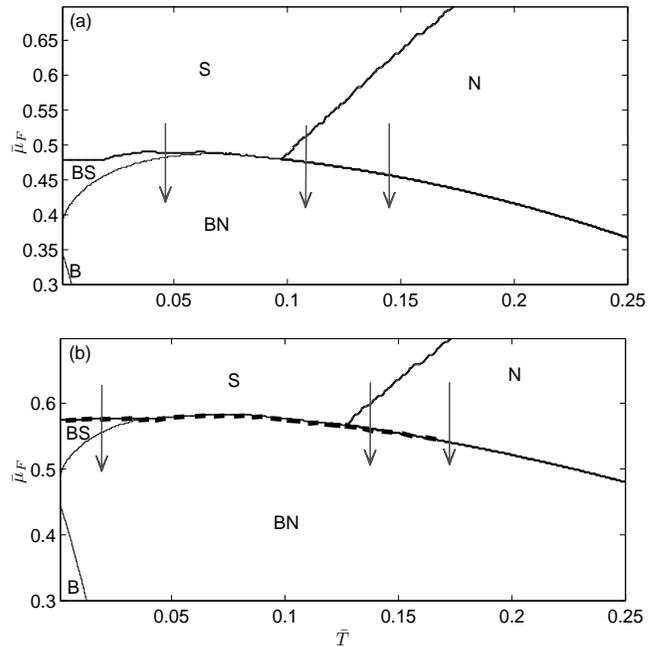}%
\caption{The phase diagram in the $T-\mu_{F}$ space when the chemical
potential for the bosons is fixed to (a) $\mu_{B}=0.35$ and (b) $\mu_{B}%
=0.45$. The features in (a) and (b) can be easily deduced from the phase
diagrams in the previous section. $U_{0}$ is same as in Fig. \ref{Fig:Zero},
and units are defined in the text. }%
\label{Fig: Tight}%
\end{center}
\end{figure}

Figure \ref{Fig: Tight} (a) displays the phase diagram for a homogeneous
system in the $T-\mu_{F}$ space for $\mu_{B}=0.35$ [slightly smaller than
$\mu_{B}^{\ast}$ in Fig. \ref{Fig:Finite}(a)]. \ \ At a given temperature and
within the local-density approximation, a potential profile $V_{F}\left(
r\right)  $ along the radially outward direction in the real space induces an
image $\mu_{F}\left(  r\right)  $ in the form of a vertical line moving down
an arrow as indicated in Fig. \ref{Fig: Tight}. \ Then, by traversing from the
tail to the head of the arrow and translating each point [$\mu_{F}\left(
r\right)  $] on the vertical line into the density via the mapping $\rho
_{F}^{h}\left(  \mu_{F}\left(  r\right)  \right)  $, we can construct, from
the center to the edge of the trap, the density distribution $\rho_{F}\left(
r\right)  \equiv\rho_{F}^{h}\left(  \mu_{F}\left(  r\right)  \right)  $, where
$\rho_{F}^{h}\left(  x\right)  $ calculates the Fermi density for a
homogeneous phase at the chemical potential $x$. \ Using the rules just
outlined and given the locations and the sizes of the vertical lines indicated
in Fig. \ref{Fig: Tight}(a), we can easily arrive at the following qualitative conclusions.

(a) At sufficiently high temperature (the arrow on the right), the density
profile consists of a normal phase (N) core surrounded by a (spherical) shell
of a BN mixture.

(b) As temperature is lowered (the arrow in the middle), superfluidity (S)
emerges from the core with N being sandwiched between S and BN.

(c) As temperature is further reduced (the arrow on the left), $N$ disappears
and substituted in its place is the Bose-superfluidity mixture (BS).

(d) A further reduction in the temperature completely eliminates $BN$,
creating a density profile, reminiscent of the high temperature case in (a),
except that $N$ and $BN$ in (a) are now replaced with $S$ and $BS$, respectively.

An intriguing aspect of such a mixture is that both critical temperatures and
Fermi profiles depend strongly on $\mu_{B}$ - the chemical potential of the
surrounding BEC. As $\mu_{B}$\ can be independently tuned in experiments, we
have at our disposal a convenient tool to selectively access the parts of
phase diagrams, that are physically interesting. \ Suppose that we want to
access the first-order phase transition. \ We can do so simply by increasing
$\mu_{B}$ beyond a certain threshold value. \ An example is given in Fig.
\ref{Fig: Tight}(b) where $\mu_{B}$ is fixed to $0.45$, a value slightly
higher than $\mu_{B}^{\ast}$ in Fig. \ref{Fig:Finite}(a). \ As can be seen,
part of the transition line dividing pure and mixed states changes its nature
from second order in Fig. \ref{Fig: Tight}(a) to first order in Fig.
\ref{Fig: Tight}(b). \ Thus, although the system can undergo a similar set of
phase transitions as in Fig. \ref{Fig: Tight}(a) [at the expense of a higher
Fermi chemical potential (or density) and a smaller BS region], a density
discontinuity is expected to emerge from Fermi profiles on a sphere that
divides the two phases sharing the same first-order transition line.%

\section{Conclusion}

\label{sec:conclusion}

A unique advantage of a cold atom system is that its system parameters can be
precisely tuned, allowing it to access a larger regime of phase diagrams.
Further, availability of detection techniques, such as absorption laser
imaging of densities and radio-frequency (RF) spectroscopy \cite{chin04},
makes the experimental determination of such phase diagrams in fine detail
possible. \ In this work, we have performed a systematic study of the
finite-temperature phase diagram in the chemical potential space for a
two-component Fermi-Bose mixture with attractive Fermi-Fermi and repulsive
Fermi-Bose interaction. Using a combination of scaling and Landau-Ginzburg
theory, we have identified, within the framework of mean-field theory, a set
of features generic to the phase diagrams for such mixtures.\ Further, we have
applied the theory to explore the physics of pairing among fermions in a
tightly confined trap surrounded by a large BEC gas.

Finally, we comment that so far, our analysis is based on the thermodynamic
potential in Eq. (\ref{Omega}) derived when all the phonon fields,
$\phi_{\mathbf{k},B}\left(  \tau\right)  $, are ignored. \ In more realistic
situations, we need to retain the phonon fields, $\phi_{\mathbf{k},B}\left(
\tau\right)  $, which, after being explicitly integrated out under the
Bogoliubov approximation, is shown to induce an attractive interaction,
$U^{ind}\left(  \mathbf{q}\right)  =-g_{BF}^{2}/g_{BB}/\left[  1+\left(
\hbar\mathbf{q}/\sqrt{4m_{B}g_{BB}n_{B}}\right)  ^{2}\right]  $, between
fermions when the retardation effect is ignored \cite{Stoof00,wang06}. \ \ The
net effect of $\phi_{\mathbf{k},B}\left(  \tau\right)  $ is then to change
$U_{0}$ to $U^{eff}\equiv U_{0}-g_{BF}^{2}/g_{BB}$, for the s-wave scattering,
which is expected to dominate all the other partial-wave scatterings. Thus,
the phase-diagram features presented in this work shall remain qualitatively
true not only for two-component Fermi-Bose mixtures with direct Fermi-Fermi
attraction ($U_{0}<0$), but also for those with direct Fermi-Fermi repulsion
$\left(  U_{0}>0\right)  $ but attractive effective Fermi-Fermi interaction
($U^{eff}<0$) \cite{viverit02}; the latter case turns out to cover several
important systems of current experimental interest, including $^{40}$K-$^{87}%
$Rb and $^{6}$Li-$^{7}$Li \cite{illuminati04}.

\section{Acknowledgement}

This work is supported by the US National Science Foundation and US Army
Research Office.

\bigskip%

\bigskip
\end{document}